\documentclass{elsart}%
\usepackage{amssymb}
\usepackage{amsfonts}
\usepackage{amsmath}
\usepackage{natbib}%
\usepackage{graphicx}
\begin{document}
\begin{frontmatter}
\title{Analytical calculation of the solid angle defined by a cylindrical detector and a point cosine source with orthogonal axes}
\author{M. J. Prata\thanksref{FCT}}
\ead{mjprata@sapo.pt}
\thanks[FCT]{Supported by Funda\c{c}\~{a}o para a Ci\^{e}ncia e Tecnologia
(Programa Praxis XXI - BD/15808/98)}
\address{Instituto  Tecnol\'{o}gico e Nuclear (ITN), Sacav\'{e}m, Portugal}
\begin{abstract}
We derive analytical expressions for the solid angle subtended by a
right circular cylinder at a point source with cosine angular distribution in the case
where the source and the cylinder axes are mutually orthogonal.
\end{abstract}
\end{frontmatter}

\section{Introduction}

The calculation of the solid angle subtended by a cylindrical detector at a
point source is of common interest in nuclear science. The case of an
isotropic source has been extensively studied, but, to the best of our
knowledge, no work has been published considering a point cosine source. Such
source could arise as the second term in a Legendre expansion of a general
source or, to take a well known example, as one of the terms in the Fermi
expression $\ \Phi(\mu)=1+\sqrt{3}\mu$. This expression holds for the angular
distribution of low energy neutrons leaking from a variety of scattering
materials and, if suitably modified ($\Phi(\mu)=1+A\mu$), can also be used in
the case of lithium hydride \citep{Verbinski}. The scattering of neutrons from
hydrogen nuclei considered at rest is also described by a cosine distribution \citep{OttBezella}.

In this work the situation of a point cosine source defined with respect to
some axis and a right circular cylinder with axis orthogonal to that of the
source is considered. Under this restriction we present the analytical
calculation of the solid angle subtended at the source positioned at an
arbitrary location. Sample graphics of the expressions obtained are presented.

\section{Solid Angle Calculation}

Let the unit vector $\mathbf{k}$ define the source direction. The source
distribution $f(\mathbf{\Omega})$ giving the probability $f(\mathbf{\Omega
})d\Omega$ of emission around the direction of the unit vector $\mathbf{\Omega
}$ is defined by%

\begin{equation}
f(\mathbf{\Omega})=\left\{
\begin{array}
[c]{cc}%
\frac{\mathbf{k}\cdot\mathbf{\Omega}}{\pi} & ,\mathbf{k}\cdot\mathbf{\Omega
\geq0}\\
0 & ,\mathbf{k}\cdot\mathbf{\Omega<0}%
\end{array}
\right.  ~.\label{f_source_def}%
\end{equation}

The distribution is normalized so that%

\begin{equation}
\iint\limits_{all\,directions}f(\mathbf{\Omega})d\Omega=1~.
\end{equation}

The solid angle ($\Omega$) is given by%

\begin{equation}
\Omega=\iint\limits_{\substack{directions \\hitting\,detector}%
}f(\mathbf{\Omega})d\Omega~.
\end{equation}

\begin{figure}[h]
\begin{center}
\includegraphics[
height=6.0319cm,
width=5.9858cm
]{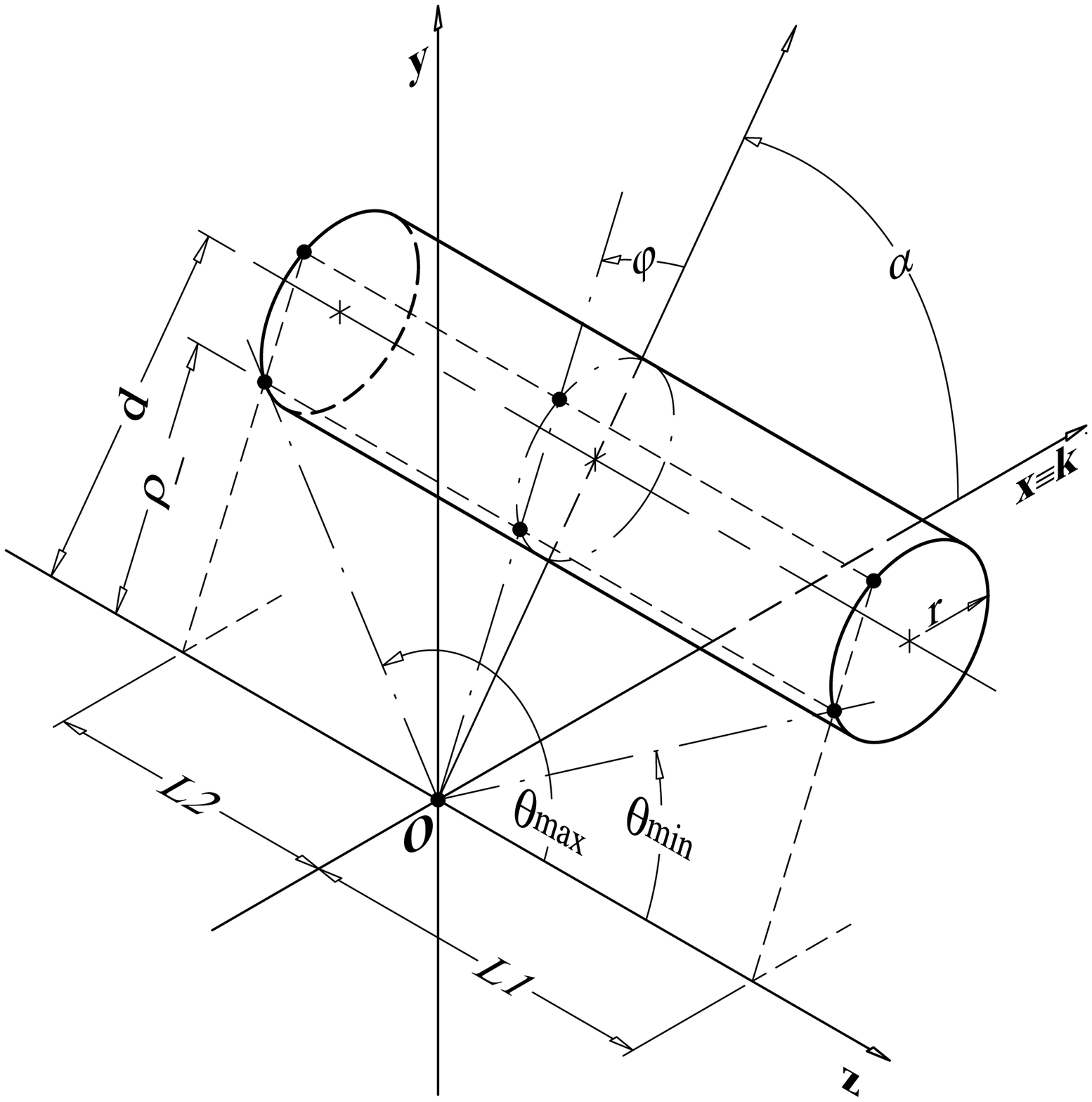}
\caption{Geometry in Case (i)}
\label{fig1}
\end{center}
\end{figure}

\begin{figure}[h]
\begin{center}
\includegraphics[
height=5.9177cm,
width=6.0912cm
]{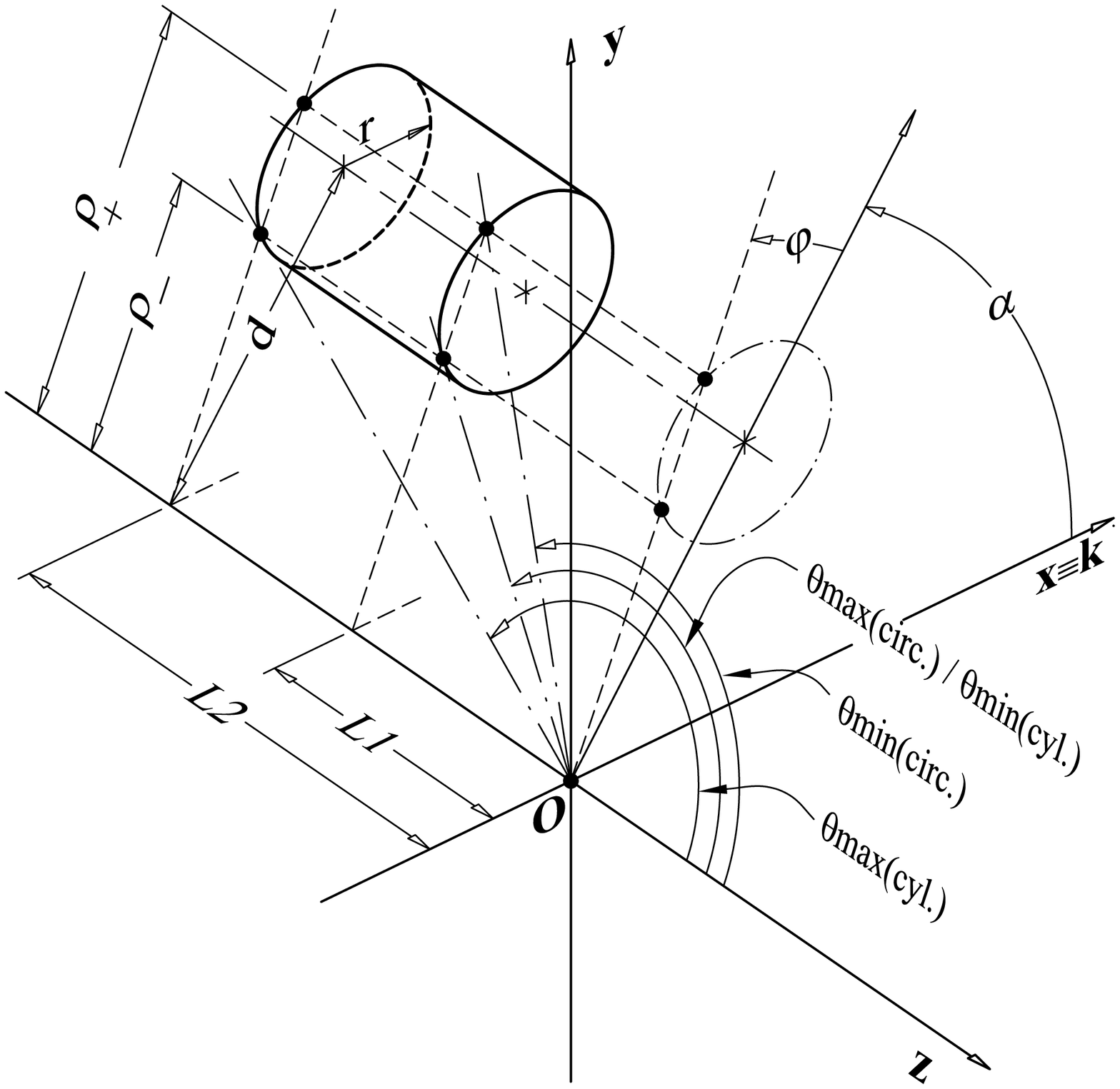}
\caption{Geometry in case (ii)}
\label{fig2}
\end{center}
\end{figure}

\begin{figure}[h]
\begin{center}
\includegraphics[
trim=0.000000in 0.000000in -0.006872in 0.000000in,
height=4.8919cm,
width=6.0934cm
]{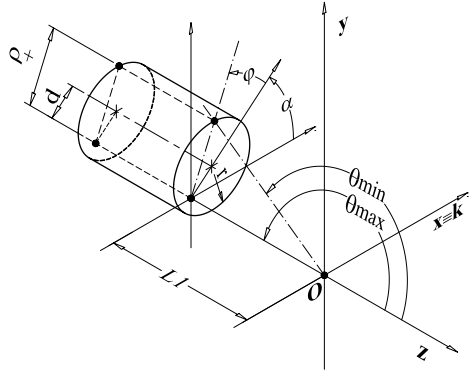}
\caption{Geometry in case (iii)}
\label{fig3}
\end{center}
\end{figure}

The origin of the coordinate system is chosen to coincide with the point
source, the $x$ axis aligned with $\mathbf{k}$ and the $z$ axis parallel to
the cylinder axis. In figs. \ref{fig1} to \ref{fig3} we show the three cases
to be considered and introduce some of the notation used. Generally the solid
angle is a sum of two terms, the one subtended by the cylindrical surface
($\Omega_{cyl}$) and the other by either of the end circles ($\Omega_{circ}$).
In case (i) one has $\Omega\equiv\Omega_{cyl}$, in case (ii) $\Omega
\equiv\Omega_{cyl}+\Omega_{circ}$ and in case (iii), $\Omega\equiv
\Omega_{circ}$. In the various cases $L_{1}$, $L_{2}$, $d$ and $r$ are all
positive. From the symmetry of the problem, the solid angle is an even
function of $\alpha$ and in the following we will thus consider $\alpha\geq0$.
In cases (i) and (ii) one has $d\geq r$ and, in case (iii), $r>d$. With the
notation used there results that%
\begin{equation}
\mathbf{k}\cdot\mathbf{\Omega}=\cos(\alpha+\varphi)\sin(\theta
)\label{cos_psi_alfa_theta}%
\end{equation}
where $\theta$ is the polar angle from the z axis and $\varphi$ is the
azimuthal angle in the $xy$ plane as measured from an axis through the origin
and intersecting the detector axis. Hence
\begin{equation}
\Omega=\frac{1}{\pi}\int\limits_{\varphi_{\min}}^{\varphi_{\max}}%
\int\limits_{\theta_{\min}}^{\theta_{\max}}\cos(\alpha+\varphi)\sin^{2}%
(\theta)d\theta d\varphi\label{Solid_angle_theta_phi}%
\end{equation}
where the limiting angles have yet to be determined.%

\subsection{Integration limits}

\begin{figure}[h]
\begin{center}
\includegraphics[
height=4.3186cm,
width=6.0934cm
]{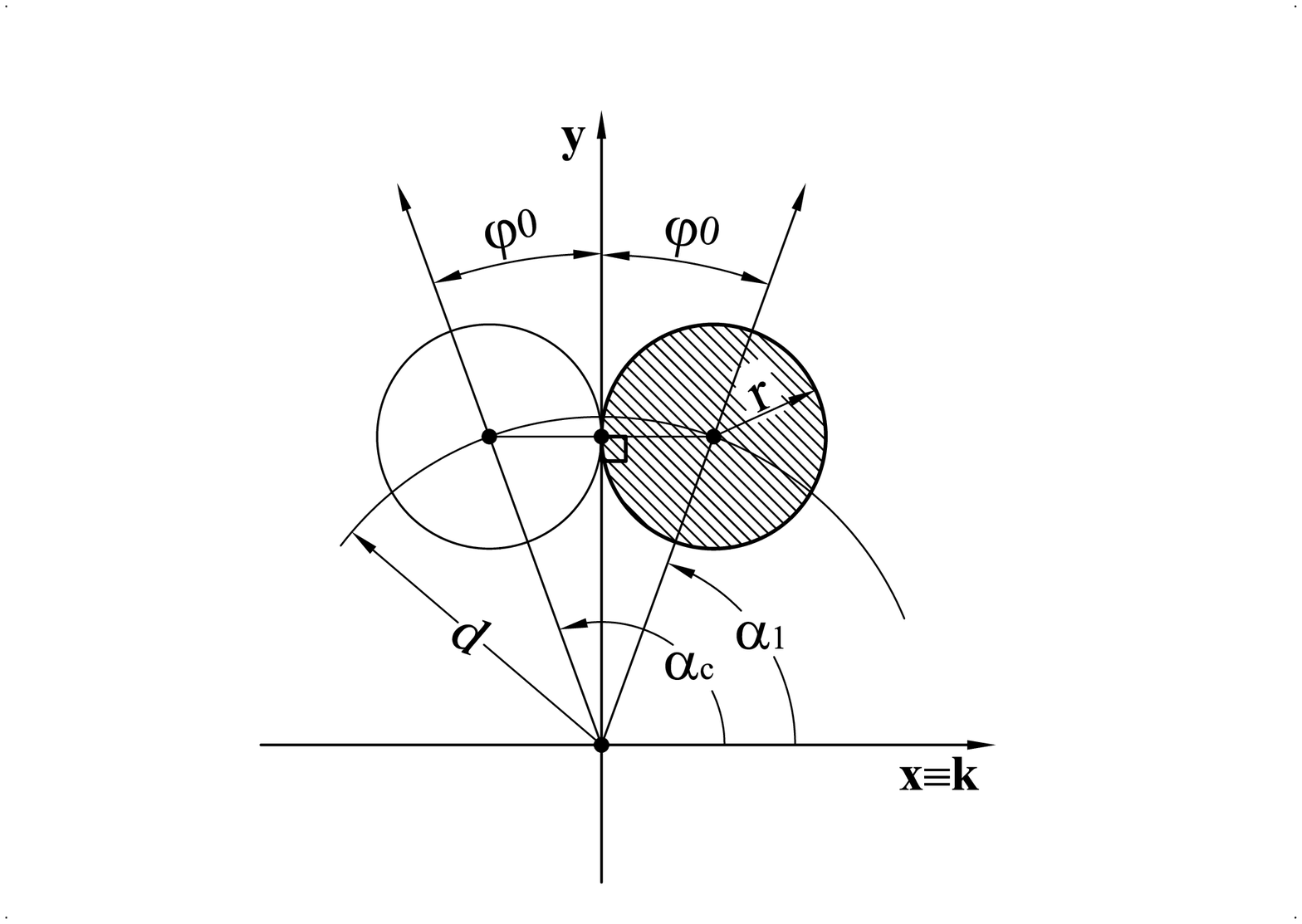}
\caption{Definitions of $\varphi_{0}$, $\alpha_{1}$ and $\alpha_{c}$ (cases
(i) and (ii) )}
\label{fig4}
\end{center}
\end{figure}

\begin{figure}[hh]
\begin{center}
\includegraphics[
height=4.3186cm,
width=6.0934cm
]{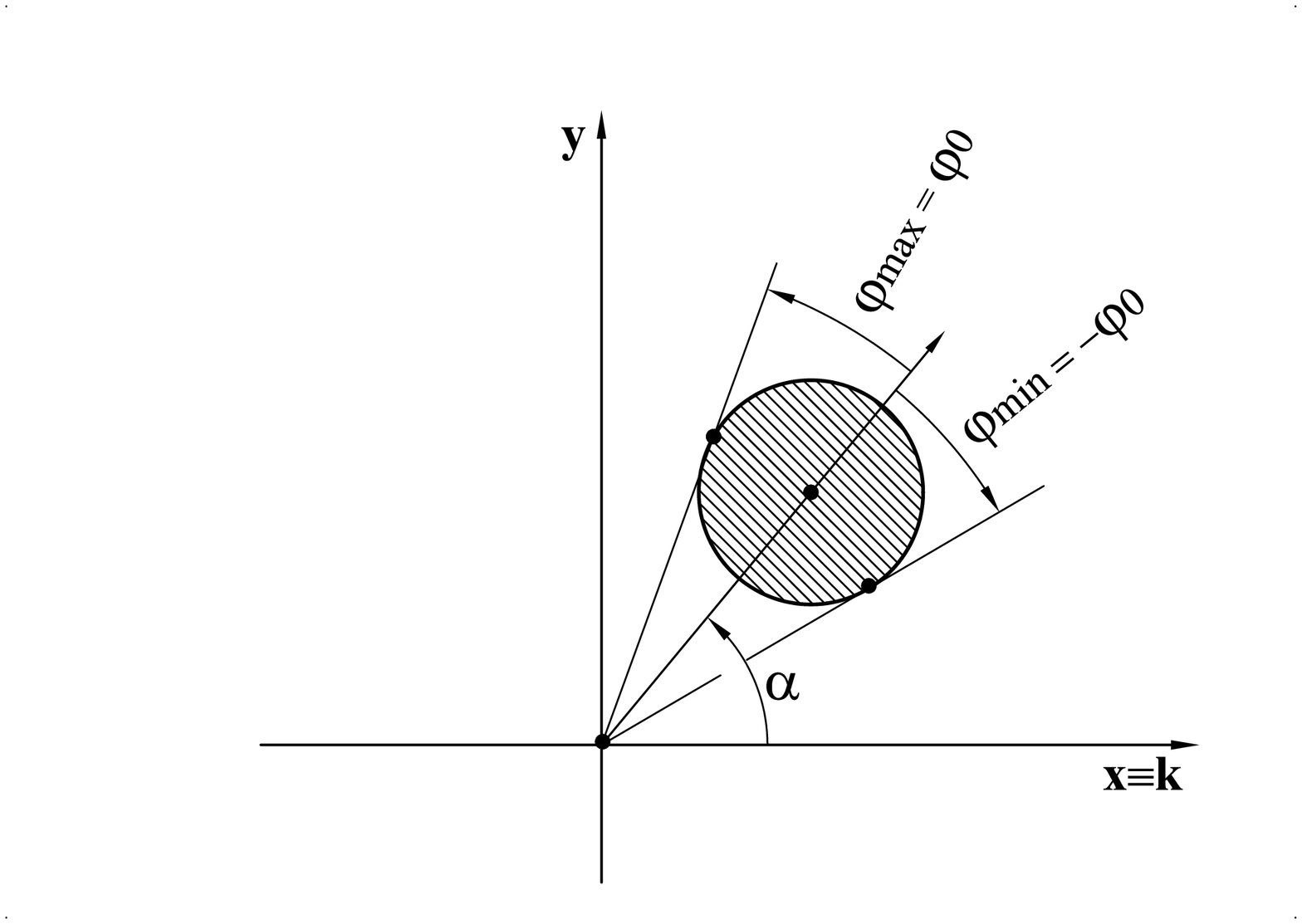}
\caption{Integration limits for $\varphi$ in cases (i) and (ii) when
$0\leq\alpha<\alpha_{1}$}
\label{fig5}
\end{center}
\end{figure}

\begin{figure}[hhptb]
\begin{center}
\includegraphics[
height=4.3164cm,
width=6.0912cm
]{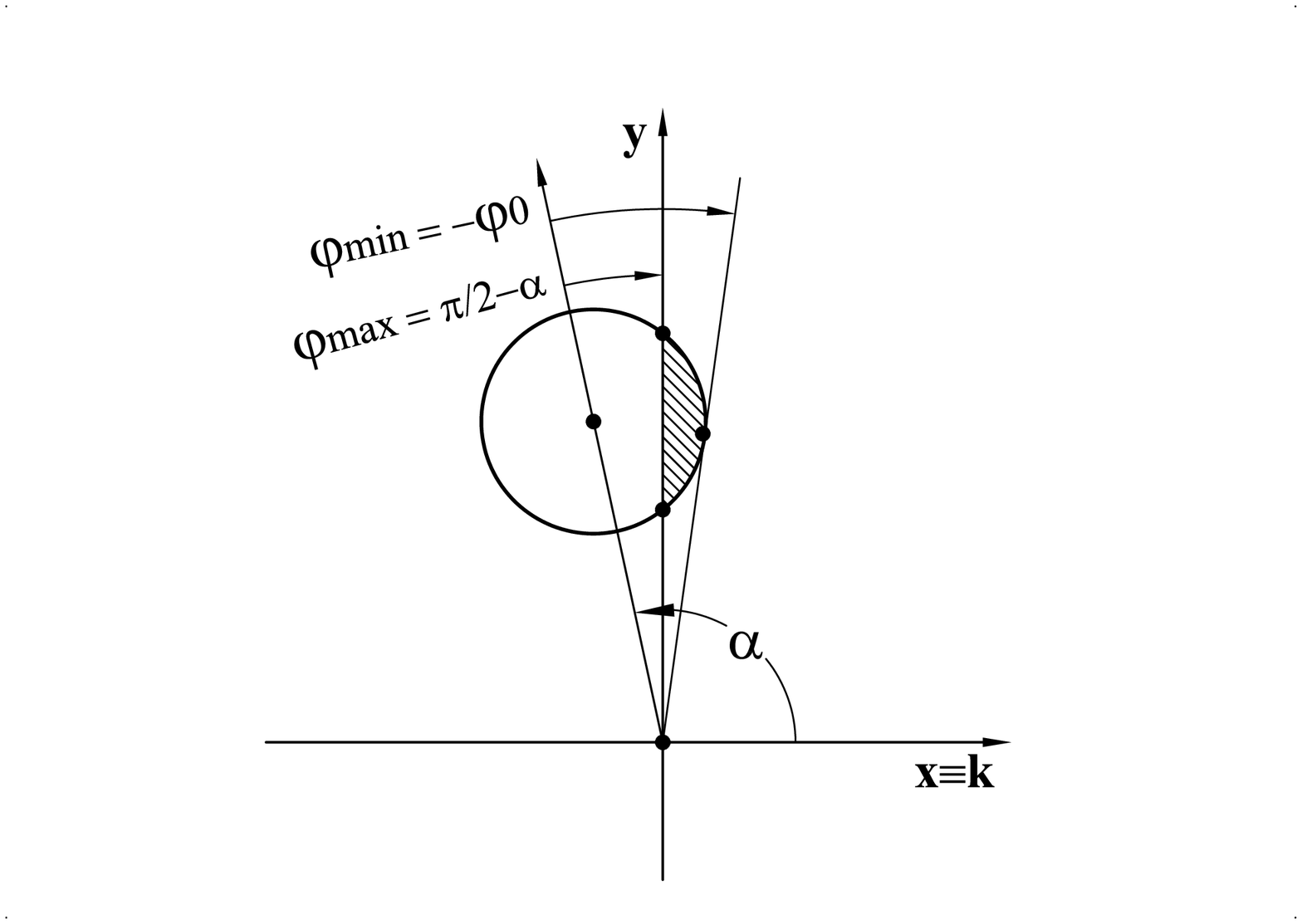}
\caption{Integration limits for $\varphi$ in cases (i) and (ii) when
$\alpha_{1}\leq\alpha<\alpha_{c}$}%
\label{fig6}
\end{center}
\end{figure}

\begin{figure}
[hhptbh]
\begin{center}
\includegraphics[
height=4.3164cm,
width=6.0912cm
]{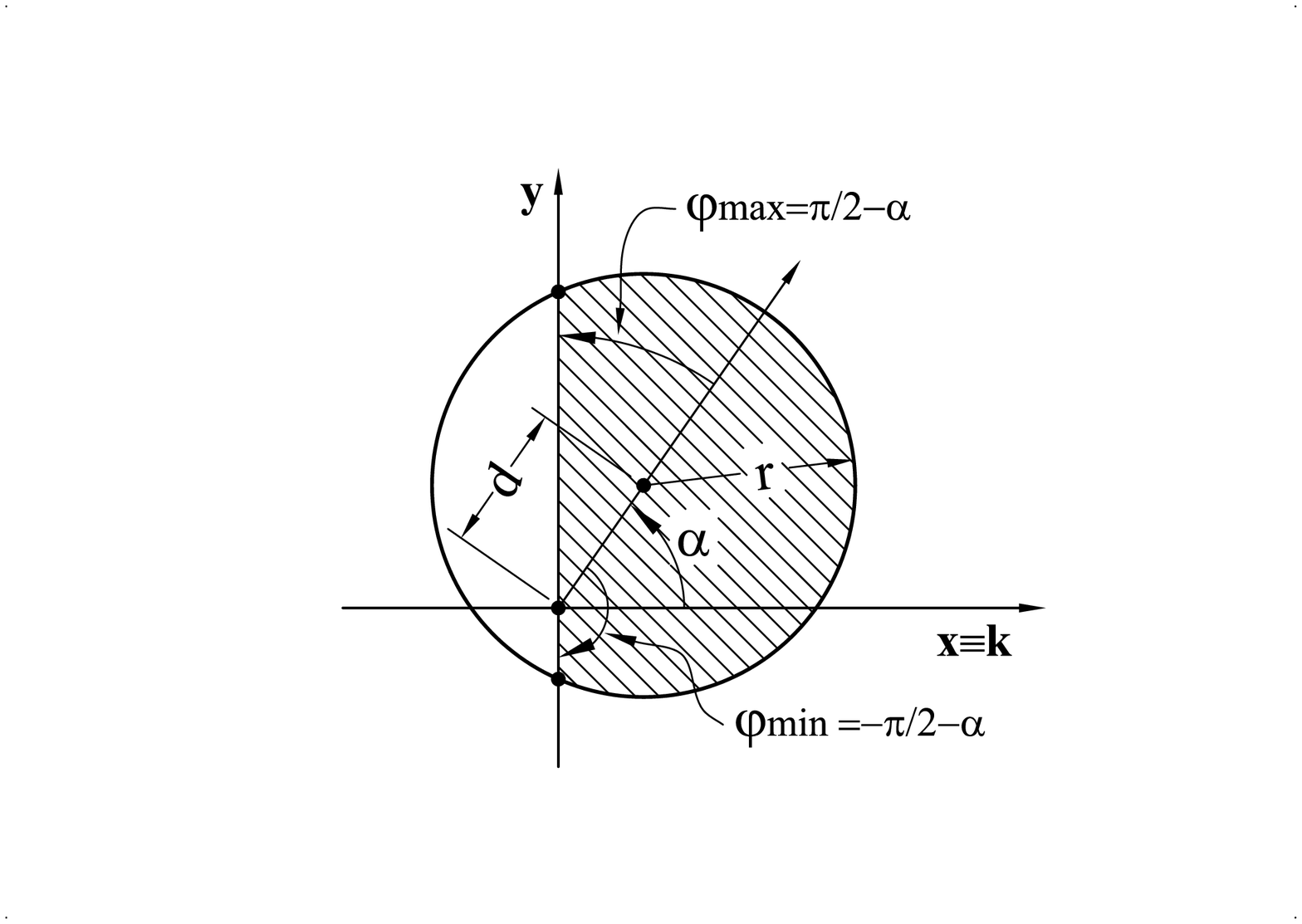}
\caption{Integration limits for $\varphi$ in case (iii)}
\label{fig7}
\end{center}
\end{figure}

Because the source only emits into the hemisphere corresponding to $x\geq0$,
the range of variation of $\varphi$ depends on the value of $\alpha$, as
explained in figs. \ref{fig4} to \ref{fig7}, where the the hashed area shows
the \textit{illuminated} part of the detector. Introducing
\begin{equation}
\varphi_{0}=\arcsin\left(  r/d\right)  ~,\label{phi0_definition}%
\end{equation}
there results%
\begin{equation}
\alpha_{1}=\pi/2-\varphi_{0}\label{alfa1_definition}%
\end{equation}
and
\begin{equation}
\alpha_{c}=\pi/2+\varphi_{0}~.\label{alfac_definition}%
\end{equation}%

The limits for $\varphi$ are then given in table \ref{tab:PhiLimits}.

\begin{table}[ptb]
\caption{Integration limits for $\varphi$}%
\label{tab:PhiLimits}%
\begin{tabular}
[c]{cccc}\hline
Case & $\alpha\,$range & $\varphi_{\min}$ & $\varphi_{\max}$\\\hline
& $0\leq\alpha<\alpha_{1}$ & $-\varphi_{0}$ & $\varphi_{0}$\\
(i) or (ii) & $\alpha_{1}\leq\alpha<\alpha_{c}$ & $-\varphi_{0}$ &
$\pi/2-\alpha$\\
& $\alpha_{c}\leq\alpha\leq\pi$ & \multicolumn{2}{c}{$\Omega=0$}\\
(iii) & $0\leq\alpha\leq\pi$ & $-\pi/2-\alpha$ & $\pi/2-\alpha$%
\end{tabular}
\end{table}

The limits for $\theta$ can be expressed in terms of $\rho_{_{-}}(\varphi) $
and $\rho_{+}(\varphi)$ that were introduced in the pictures and are given by%

\begin{equation}
\rho\pm(\varphi)=d\cos(\varphi)\pm\sqrt{r^{2}-(d\sin(\varphi))^{2}%
}~.\label{rho_plus_minus_def}%
\end{equation}%

\begin{figure}
[ptb]
\begin{center}
\includegraphics[
height=4.3186cm,
width=6.0934cm
]{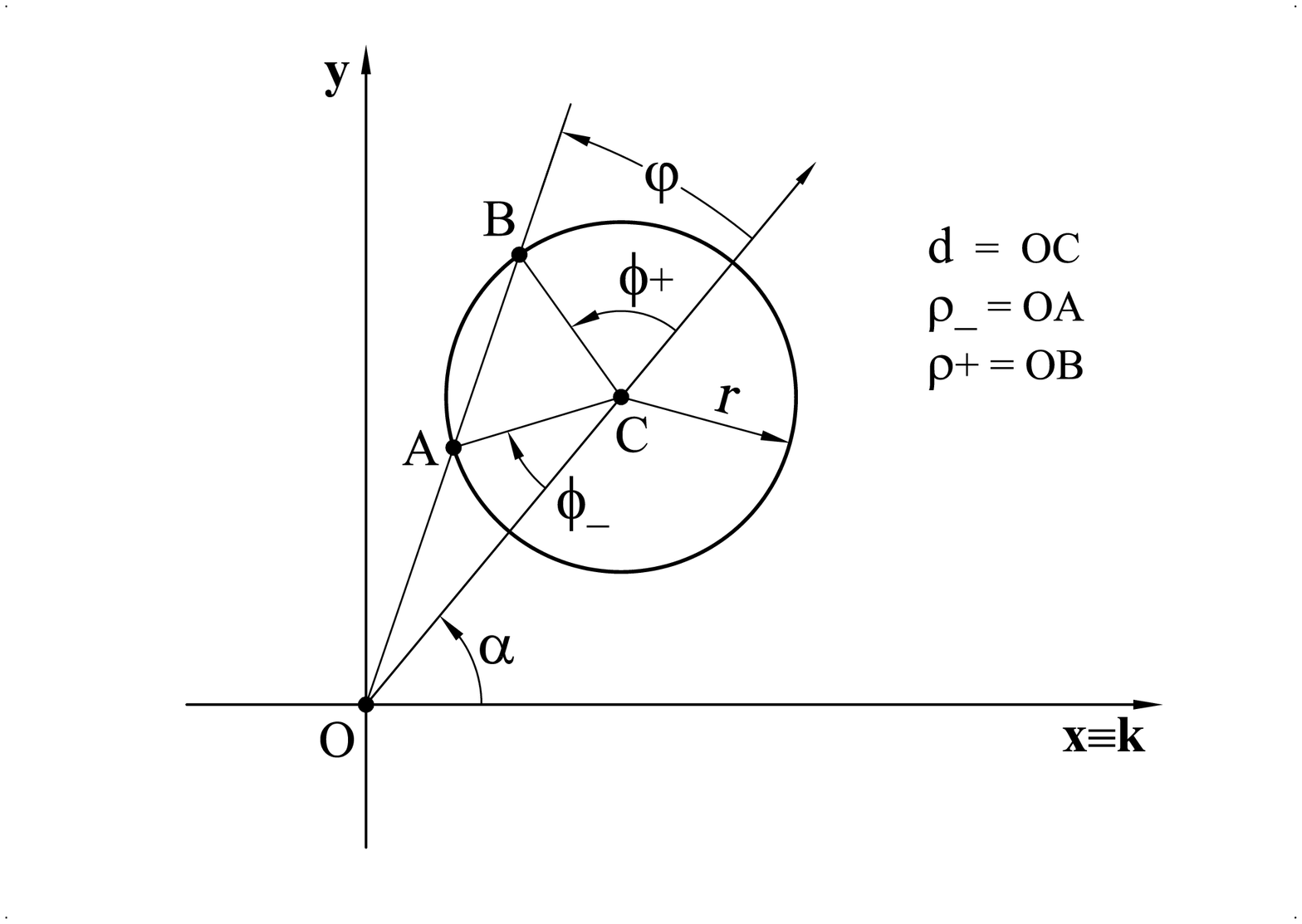}
\caption{Angles $\phi\pm$ in cases (i) and (ii) ($d\geqslant r$ )}
\label{fig8}
\end{center}
\end{figure}

\begin{figure}
[ptbptb]
\begin{center}
\includegraphics[
height=4.3186cm,
width=6.0934cm
]{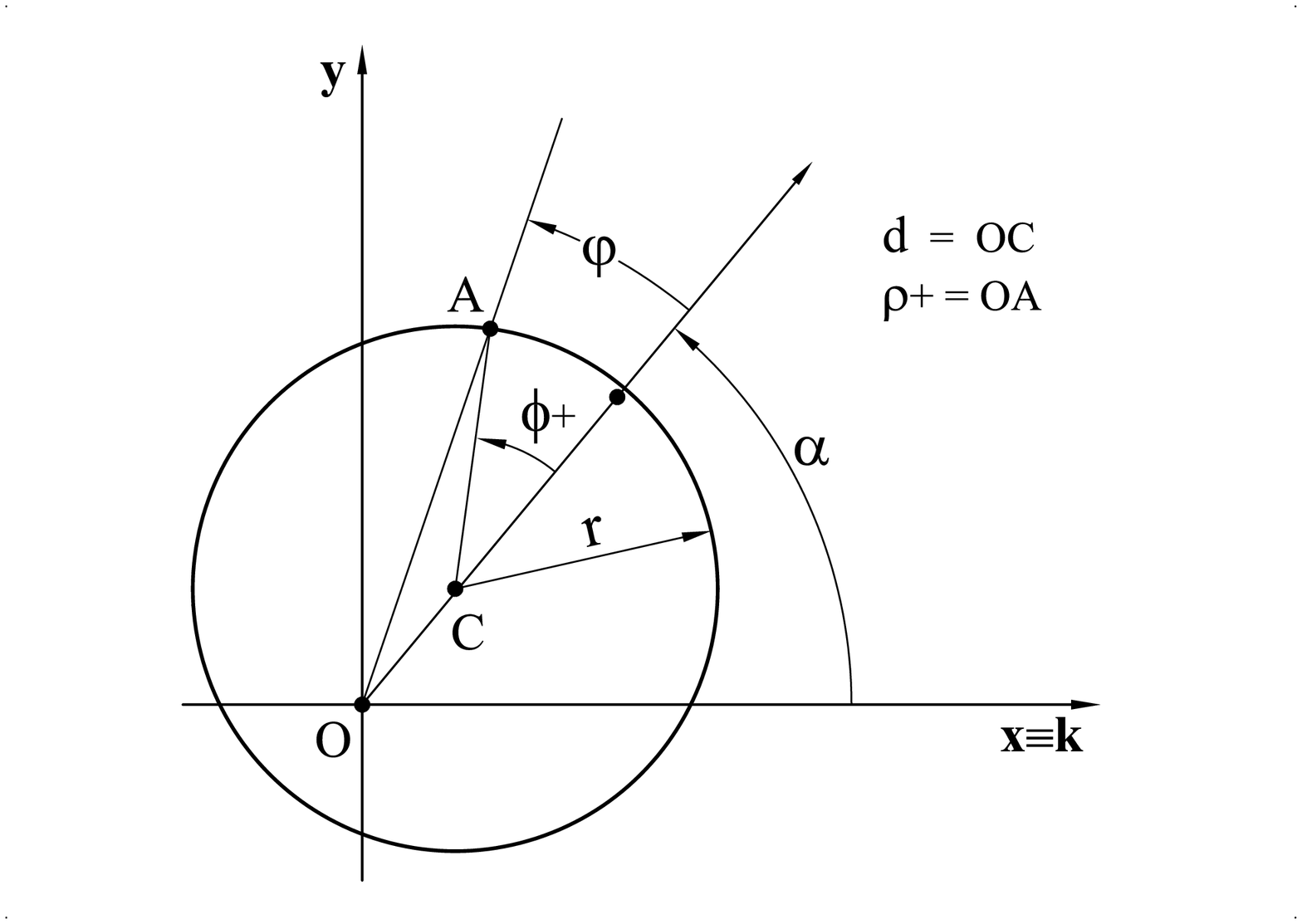}
\caption{Angle $\phi_{+}$ in case (iii) ($d<r$)}
\label{fig9}
\end{center}
\end{figure}

The limits for $\theta$ are summarized in table \ref{tab:ThetaLimits}.

\begin{table}[ptb]
\caption{Integration limits for $\theta$}%
\label{tab:ThetaLimits}%
\begin{tabular}
[c]{cccc}\hline
Case & $\Omega_{cyl}$ or $\Omega_{circ}$ & $\theta_{\min}$ & $\theta_{\max}%
$\\\hline
(i) & $\Omega_{cyl}$ & $\pi/2-\arctan(\frac{L_{1}}{\rho_{-}})$ &
$\pi/2+\arctan(\frac{L_{2}}{\rho_{-}})$\\
(ii) & $\Omega_{cyl}$ & $\pi/2+\arctan(\frac{L_{1}}{\rho_{-}})$ &
$\pi/2+\arctan(\frac{L_{2}}{\rho_{-}})$\\
(ii) & $\Omega_{circ}$ & $\pi/2+\arctan(\frac{L_{1}}{\rho_{+}})$ &
$\pi/2+\arctan(\frac{L_{1}}{\rho_{-}})$\\
(iii) & $\Omega_{circ}$ & $\pi/2+\arctan(\frac{L_{1}}{\rho_{+}})$ & $\pi$%
\end{tabular}
\end{table}

\subsection{Integration}

Let us define%

\begin{equation}
f_{1}(L,d,r,\alpha,\varphi)\equiv f_{1}(\varphi)=\frac{1}{2\pi}\int\cos
(\alpha+\varphi)\left[  \arctan\left(  \frac{L}{\rho_{_{-}}}\right)
+\frac{L\,\rho_{_{-}}}{L^{2}+\rho_{_{-}}^{2}}\right]  d\varphi\label{f1_def}%
\end{equation}

and%

\begin{equation}
f_{2}(L,d,r,\alpha,\varphi)\equiv f_{2}(\varphi)=\frac{1}{2\pi}\int\cos
(\alpha+\varphi)\left[  \arctan\left(  \frac{L}{\rho_{+}}\right)
+\frac{L\,\rho_{+}}{L^{2}+\rho_{_{+}}^{2}}\right]  d\varphi\label{f2_def}%
\end{equation}

Then, performing the first integration in the rhs of eq.
\ref{Solid_angle_theta_phi}, where $\theta_{\min}$ and $\theta_{\max}$ are
given in table \ref{tab:ThetaLimits}\textbf{, }yields in each case:

case (i)%

\begin{equation}
\Omega=\Omega_{cyl}=\left[  f_{1}(L_{1},d,r,\alpha,\varphi)+f_{1}%
(L_{2},d,r,\alpha,\varphi)\right]  _{\varphi_{\min}}^{\varphi_{\max}%
}~,\label{omega_f_case1}%
\end{equation}

case (ii)%

\begin{equation}
\Omega_{cyl}=\left[  f_{1}(L_{2},d,r,\alpha,\varphi)-f_{1}(L_{1}%
,d,r,\alpha,\varphi)\right]  _{\varphi_{\min}}^{\varphi_{\max}}%
~,\label{omega_f_case2_cyl}%
\end{equation}%
\begin{equation}
\Omega_{circ}=\left[  f_{1}(L_{1},d,r,\alpha,\varphi)-f_{2}(L_{1}%
,d,r,\alpha,\varphi)\right]  _{\varphi_{\min}}^{\varphi_{\max}}%
~,\label{omega_f_case2_circ}%
\end{equation}

\begin{equation}
\Omega=\Omega_{cyl}+\Omega_{circ}=\left[  f_{1}(L_{2},d,r,\alpha
,\varphi)-f_{2}(L_{1},d,r,\alpha,\varphi)\right]  _{\varphi_{\min}}%
^{\varphi_{\max}}~,\label{omega_f_case2}%
\end{equation}

case (iii)%

\begin{align}
\Omega & =\Omega_{circ}=\frac{1}{4}\int_{\varphi_{\min}}^{\varphi_{\max}%
}\mathbf{\cos}(\alpha+\varphi)d\varphi-f_{2}(L_{1},d,r,\alpha,\varphi
)|_{\varphi_{\min}}^{\varphi_{\max}}~,\label{omega_f_case3_no_simplification}%
\\
& =\frac{1}{2}-f_{2}(L_{1},d,r,\alpha,\varphi)|_{\varphi_{\min}}%
^{\varphi_{\max}}~.\label{omega_f_case3}%
\end{align}

Eq. \ref{omega_f_case3} is obtained from eq.
\ref{omega_f_case3_no_simplification} by integration of the 1st term using the
values from table \ref{tab:PhiLimits}.

Using again the limits from table \ref{tab:PhiLimits}, eqs. \ref{omega_f_case1} to
\ref{omega_f_case2} and \ref{omega_f_case3} \ can be rewritten as

case (i)%

\begin{equation}
\Omega=\Omega_{cyl}=F_{1}(L_{1},d,r,\alpha)+F_{1}(L_{2},d,r,\alpha
)~,\label{omega_F_case1}%
\end{equation}

case (ii)%

\begin{equation}
\Omega_{cyl}=F_{1}(L_{2},d,r,\alpha)-F_{1}(L_{1},d,r,\alpha)~,
\end{equation}%
\begin{equation}
\Omega_{circ}=F_{1}(L_{1},d,r,\alpha)-F_{2}(L_{1},d,r,\alpha)~,
\end{equation}

\begin{equation}
\Omega=\Omega_{cyl}+\Omega_{circ}=F_{1}(L_{2},d,r,\alpha)-F_{2}(L_{1}%
,d,r,\alpha)~,\label{omega_F_case2}%
\end{equation}

case (iii)%

\begin{equation}
\Omega=\Omega_{circ}=F_{3}(L_{1},d,r,\alpha)~,\label{omega_F_case3}%
\end{equation}

where the integrals $F_{i}\;(i=1..3)$ are defined in table
\ref{tab:IntegralsFiDef}.

\bigskip\begin{table}[ptb]
\caption{Definition of the integrals $F_{i}$}%
\label{tab:IntegralsFiDef}%
\begin{tabular}
[c]{ccccc}\hline
Integral & $\alpha\,$range & Value & $\varphi_{\min}$ & $\varphi_{\max}%
$\\\hline
& $0\leq\alpha<\alpha_{1}$ & $f_{i}(L,d,r,\alpha,\varphi)|_{\varphi_{\min}%
}^{\varphi_{\max}}$ & $-\varphi_{0}$ & $\varphi_{0}$\\
$F_{i}$ \ $(i=1,2)$ & $\alpha_{1}\leq\alpha<\alpha_{c}$ & $f_{i}%
(L,d,r,\alpha,\varphi)|_{\varphi_{\min}}^{\varphi_{\max}}$ & $-\varphi_{0}$ &
$\pi/2-\alpha$\\
& $\alpha_{c}\leq\alpha\leq\pi$ & $0$ & - & -\\
$F_{3}$ & $0\leq\alpha\leq\pi$ & $1/2-f_{2}(L,d,r,\alpha,\varphi
)|_{\varphi_{\min}}^{\varphi_{\max}}$ & $-\pi/2-\alpha$ & $\pi/2-\alpha$%
\end{tabular}
\end{table}

\subsubsection{Calculation of $F_{1}$}

Integrating by parts the 1st term in the rhs of eq. \ref{f1_def},%

\begin{gather}
\int\cos(\alpha+\varphi)\arctan\left(  \frac{L}{\rho_{_{-}}}\right)
d\varphi=\sin(\alpha+\varphi)\arctan\left(  \frac{L}{\rho_{_{-}}}\right)
\nonumber\\
+\int\sin(\alpha+\varphi)\frac{L}{L^{2}+\rho_{_{-}}^{2}}\frac{d\sin
(\varphi)\rho_{_{-}}}{\sqrt{r^{2}-\left[  d\sin(\varphi)\right]  ^{2}}%
}d\varphi\label{by_parts_f1_bis_}%
\end{gather}

Substituting eq. \ref{by_parts_f1_bis_} in the rhs of eq. \ref{f1_def}
\ yields, after some algebra,%
\begin{equation}
f_{1}(\varphi)=\frac{1}{2\pi}\left[  A_{10}+\cos(\alpha)A_{11}+\sin
(\alpha)A_{12}\right]  ~,\label{f1_A0_A1_A2}%
\end{equation}

where%

\begin{equation}
A_{10}=\sin(\alpha+\varphi)\arctan\left(  \frac{L}{\rho_{_{-}}}\right)
~,\label{A10_def}%
\end{equation}

\begin{equation}
A_{11}=\int\frac{L\rho_{_{-}}}{L^{2}+\rho_{_{-}}^{2}}\left(  \cos
(\varphi)+\frac{d\sin^{2}(\varphi)}{\sqrt{r^{2}-\left[  d\sin(\varphi)\right]
^{2}}}\right)  d\varphi\label{A11_def}%
\end{equation}

and%

\begin{equation}
A_{12}=\int\frac{L\rho_{_{-}}}{L^{2}+\rho_{_{-}}^{2}}\left(  \frac{\rho_{_{-}%
}\sin(\varphi)}{\sqrt{r^{2}-\left[  d\sin(\varphi)\right]  ^{2}}}\right)
d\varphi~.\label{A12_def}%
\end{equation}

Performing a change of integration variable to $\phi_{-}$ represented in fig.
\ref{fig8},%

\begin{equation}
\phi_{-}=2\arctan\left(  \frac{\rho_{_{-}}\sin(\varphi)}{d+r-\rho_{_{-}}%
\cos(\varphi)}\right)  ~,\label{phi_minus_def}%
\end{equation}

we find%

\begin{equation}
A_{11}=Lr\int\frac{\cos(\phi_{-})}{L^{2}+\rho_{-}^{2}\left(  \phi_{-}\right)
}d\phi_{-}%
\end{equation}

and%

\begin{equation}
A_{12}=Lr\int\frac{\sin(\phi_{-})}{L^{2}+\rho_{-}^{2}\left(  \phi_{-}\right)
}d\phi_{-}~,
\end{equation}

where%

\begin{equation}
\rho_{_{-}}(\phi_{-})=\sqrt{d^{2}+r^{2}-2dr\cos(\phi_{-})}~.
\end{equation}

One then easily obtains%

\begin{equation}
A_{11}=\frac{L}{2d}\left[  \frac{2}{\sqrt{1-m^{2}}}\arctan\left(  \sqrt
{\frac{1+m}{1-m}}\tan\left(  \frac{\phi_{-}}{2}\right)  \right)  -\phi
_{-}\right]  ~,\label{A11_integrate}%
\end{equation}

where%

\begin{equation}
m=\frac{2dr}{L^{2}+d^{2}+r^{2}}\;\;;~0<m<1\label{m_def}%
\end{equation}

and%

\begin{equation}
A_{12}=\frac{L}{2d}\log(L^{2}+\rho_{_{-}}^{2})~.\label{A12_integrate}%
\end{equation}

\bigskip

\subsubsection{Evaluation of $F_{1}$ ($d\neq r$)\label{F1_evaluation}}

For $0\leq\alpha<\alpha_{1}$ the terms proportional to $\sin(\alpha)$ vanish
and one gets%

\begin{gather}
F_{1}(L,d,r,\alpha)=\frac{\cos(\alpha)}{\pi}\left[  \frac{r}{d}\arctan\left(
\frac{L}{\sqrt{d^{2}-r^{2}}}\right)  \right. \nonumber\\
\left.  +\frac{L}{d}\left(  \frac{1}{\sqrt{1-m^{2}}}\arctan\left(  \sqrt
{\frac{1+m}{1-m}}\sqrt{\frac{d-r}{d+r}}\right)  -\arctan\left(  \sqrt
{\frac{d-r}{d+r}}\right)  \right)  \right]  .\label{F1_even}%
\end{gather}

\bigskip

For $\ \alpha_{1}\leq\alpha<\alpha_{c}$, the expressions are less simple and
we give each term in eq. \ref{f1_A0_A1_A2}\ separately. Setting%

\begin{align}
\widetilde{\rho}_{-}  & =\rho_{-}(\varphi_{\max})=\rho_{_{-}}|_{\varphi
=\pi/2-\alpha}\\
& =d\sin\left(  \alpha\right)  -\sqrt{r^{2}-\left[  d\cos\left(
\alpha\right)  \right]  ^{2}}~,\label{rho_minus_tilde_def}%
\end{align}
the first term can be written as%

\begin{equation}
A_{10}|_{\varphi_{\min}}^{\varphi_{\max}}=\arctan\left(  \frac{L}%
{\widetilde{\rho}_{-}}\right)  -\arctan\left(  \frac{L}{\sqrt{d^{2}-r^{2}}%
}\right)  \sin(\alpha-\varphi_{0})~.\label{A10_evaluation}%
\end{equation}
Defining%
\begin{equation}
a_{1}=\tan\left[  \frac{\phi_{-}}{2}\right]  _{\varphi_{\min}}=-\sqrt
{\frac{d-r}{d+r}}%
\end{equation}
and%

\begin{equation}
b_{1}=\tan\left[  \frac{\phi_{-}}{2}\right]  _{\varphi_{\max}}=\frac
{\cos\left(  \alpha\right)  \widetilde{\rho}_{-}}{r+d-\sin\left(
\alpha\right)  \widetilde{\rho}_{-}}~,
\end{equation}
then%

\begin{gather}
A_{11}|_{\varphi_{\min}}^{\varphi_{\max}}=\frac{L}{2d}\left\{  \frac{2}%
{\sqrt{1-m^{2}}}\left[  \arctan\left(  \sqrt{\frac{1+m}{1-m}}b_{1}\right)
-\arctan\left(  \sqrt{\frac{1+m}{1-m}}a_{1}\right)  \right]  \right.
\nonumber\\
\left.  -2\left[  \arctan\left(  b_{1}\right)  -\arctan\left(  a_{1}\right)
\right]  \right\} \label{A11_evaluation}%
\end{gather}
Finally, from $A_{12}$ results%

\begin{equation}
A_{12}|_{\varphi_{\min}}^{\varphi_{\max}}=\frac{L}{2d}\log[\frac
{L^{2}+\widetilde{\rho}_{-}^{2}}{L^{2}+d^{2}-r^{2}}]~.\label{A12_evaluation}%
\end{equation}
Using eqs. \ref{f1_A0_A1_A2}, \ref{A10_evaluation}, \ref{A11_evaluation} and
\ref{A12_evaluation}, $F_{1}$ can be calculated in the case $\alpha_{1}%
\leq\alpha<\alpha_{c}$, $d\neq r$.

\subsubsection{Evaluation of $F_{1}$ ($d=r$)}

The value of $F_{1}$ when $d=r$ can be obtained from eq. \ref{f1_def} using
the integration limits from table \ref{tab:IntegralsFiDef} or calculating the
limit of the expressions obtained in \ref{F1_evaluation}, for $\alpha
\geqslant\alpha_{1}$. When $L\neq0$ one obtains $F_{1}(L,r,r,\alpha
)=(1+\cos\alpha)/4$. Since $F_{1}=0$ for $L=0$, $F_{1}$ is discontinuous when
$r=d$, $L\rightarrow0$.

\subsubsection{\bigskip Calculation of $F_{2}$}

The calculation is very similar to that of $F_{1}$. Integration by parts gives%

\begin{gather}
\int\cos(\alpha+\varphi)\arctan\left(  \frac{L}{\rho_{+}}\right)
d\varphi=\sin(\alpha+\varphi)\arctan\left(  \frac{L}{\rho_{+}}\right)
\nonumber\\
-\int\sin(\alpha+\varphi)\frac{L}{L^{2}+\rho_{+}^{2}}\frac{d\sin(\varphi
)\rho_{+}}{\sqrt{r^{2}-\left[  d\sin(\varphi)\right]  ^{2}}}d\varphi
~.\label{by_parts_f2}%
\end{gather}
Eq\textbf{. }\ref{f2_def}\textbf{ }can then be rewritten as%

\begin{equation}
f2(\varphi)=\frac{1}{2\pi}\left[  A_{20}+\cos(\alpha)A_{21}+\sin(\alpha
)A_{22}\right]  ~,\label{f2_A0_A1_A2}%
\end{equation}
where%
\begin{equation}
A_{20}=\sin(\alpha+\varphi)\arctan\left(  \frac{L}{\rho_{+}}\right)
~,\label{A20_def}%
\end{equation}

\begin{equation}
A_{21}=\int\frac{L\rho_{+}}{L^{2}+\rho_{+}^{2}}\left(  \cos(\varphi
)-\frac{d\sin^{2}(\varphi)}{\sqrt{r^{2}-\left[  d\sin(\varphi)\right]  ^{2}}%
}\right)  d\varphi\label{A21_def}%
\end{equation}

and%

\begin{equation}
A_{22}=-\int\frac{L\rho_{+}}{L^{2}+\rho_{+}^{2}}\left(  \frac{\rho_{+}%
\sin(\varphi)}{\sqrt{r^{2}-\left[  d\sin(\varphi)\right]  ^{2}}}\right)
d\varphi~.\label{A22_def}%
\end{equation}
The integration variable is now changed to $\phi_{+}$ (fig. \ref{fig8})%

\begin{equation}
\phi_{+}=2\arctan\left(  \frac{\rho_{+}\sin(\varphi)}{r-d+\rho_{+}\cos
(\varphi)}\right) \label{phi_plus_def}%
\end{equation}
and the integrals are then expressed as%

\begin{equation}
A_{21}=Lr\int\frac{\cos(\phi_{+})}{L^{2}+\rho_{+}^{2}\left(  \phi_{+}\right)
}d\phi_{+}~,
\end{equation}

\begin{equation}
A_{22}=-Lr\int\frac{\sin(\phi_{+})}{L^{2}+\rho_{+}^{2}\left(  \phi_{+}\right)
}d\phi_{+}~,
\end{equation}
where%

\begin{equation}
\rho_{+}(\phi_{+})=\sqrt{d^{2}+r^{2}+2dr\cos(\phi_{+})}~.
\end{equation}
The integration is straightforward resulting in%

\begin{equation}
A_{21}=\frac{L}{2d}\left[  \phi_{+}-\frac{2}{\sqrt{1-m^{2}}}\arctan\left(
\sqrt{\frac{1-m}{1+m}}\tan\left(  \frac{\phi_{+}}{2}\right)  \right)  \right]
\label{A21_integrate}%
\end{equation}
and%
\begin{equation}
A_{22}=\frac{L}{2d}\log(L^{2}+\rho_{+}^{2})~.\label{A22_integrate}%
\end{equation}
where $m$ is obtained from eq. \ref{m_def}.

\subsubsection{Evaluation of $F_{2}$ ($d\neq r$)\label{F2_evaluation}}

For $0\leq\alpha<\alpha_{1}$ the terms proportional to $\sin(\alpha)$ again
vanish giving%

\begin{gather}
F_{2}(L,d,r,\alpha)=\frac{\cos(\alpha)}{\pi}\left[  \frac{r}{d}\arctan\left(
\frac{L}{\sqrt{d^{2}-r^{2}}}\right)  \right. \nonumber\\
\left.  -\frac{L}{d}\left(  \frac{1}{\sqrt{1-m^{2}}}\arctan\left(  \sqrt
{\frac{1-m}{1+m}}\sqrt{\frac{d+r}{d-r}}\right)  -\arctan\left(  \sqrt
{\frac{d+r}{d-r}}\right)  \right)  \right]  ~.\label{F2_even}%
\end{gather}
It is interesting that using $\arctan(z)+\arctan(1/z)=\pi/2$ the expression
can be recast as%

\begin{equation}
F_{2}(L,d,r,\alpha)=F_{1}(L,d,r,\alpha)+\cos(\alpha)\,L/(2d)\,\left(
1-1/\sqrt{1-m^{2}}\right)  ~.\label{F2_even_eq_F!_even_plus_remainder}%
\end{equation}
For $\ \alpha_{1}\leq\alpha<\alpha_{c}$, setting%

\begin{align}
\widetilde{\rho}_{+}  & =\rho_{+}(\varphi_{\max})=\rho_{+}|_{\varphi
=\pi/2-\alpha}\\
& =d\sin\left(  \alpha\right)  +\sqrt{r^{2}-\left(  d\cos\left(
\alpha\right)  \right)  ^{2}}~,\label{rho_plus_tilde_def}%
\end{align}
the first term is determined from%

\begin{equation}
A_{20}|_{\varphi_{\min}}^{\varphi_{\max}}=\arctan\left(  \frac{L}%
{\widetilde{\rho}_{+}}\right)  -\arctan\left(  \frac{L}{\sqrt{d^{2}-r^{2}}%
}\right)  \sin(\alpha-\varphi_{0})~.\label{A20_evaluation}%
\end{equation}
Considering%
\begin{align}
a_{2}  & =\tan\left[  \frac{\phi_{+}}{2}\right]  _{\varphi_{\min}}%
=-\sqrt{\frac{d+r}{d-r}}~,\\
b_{2}  & =\tan\left[  \frac{\phi_{+}}{2}\right]  _{\varphi_{\max}}=\frac
{\cos\left(  \alpha\right)  \widetilde{\rho}_{+}}{r-d+\sin\left(
\alpha\right)  \widetilde{\rho}_{+}}~,
\end{align}
gives%

\begin{gather}
A_{21}|_{\varphi_{\min}}^{\varphi_{\max}}=\frac{L}{2d}\left\{  2\left[
\arctan\left(  b_{2}\right)  -\arctan\left(  a_{2}\right)  \right]  \right.
\nonumber\\
\left.  -\frac{2}{\sqrt{1-m^{2}}}\left[  \arctan\left(  \sqrt{\frac{1-m}{1+m}%
}b_{2}\right)  -\arctan\left(  \sqrt{\frac{1-m}{1+m}}a_{2}\right)  \right]
\right\}  ~.\label{A21_evaluation}%
\end{gather}
\linebreak Finally, for $A_{22}$ the result is%

\begin{equation}
A_{22}|_{\varphi_{\min}}^{\varphi_{\max}}=\frac{L}{2d}\log[\frac
{L^{2}+\widetilde{\rho}_{+}^{2}}{L^{2}+d^{2}-r^{2}}]~,\label{A22_evaluation}%
\end{equation}
which completes the evaluation of $F_{2}$ when $r\neq d$.

\subsubsection{Evaluation of $F_{2}$ ($d=r$)}

Taking the limit of the expressions obtained in \ref{F2_evaluation} when
$L\neq0$, one obtains $F_{2}(L,r,r,\alpha=0)=1/2+L/(2r)(1-1/\sqrt{1-m^{2}})$,
$F_{2}(L,r,r,\alpha=\pi)=0$ and%

\begin{gather}
F_{2}(L,r,r,\alpha)=(1+\cos\alpha)/4-1/(2\pi)\{\arctan\left[  2(r/L)\sin
\alpha\right] \nonumber\\
+\cos(\alpha)L/r\left(  \alpha-\pi+1/\sqrt{1-m^{2}}\left(  \pi/2+\arctan
\left[  \cot(\alpha)\sqrt{1-m}/\sqrt{1+m}\right]  \right)  \right) \nonumber\\
-\sin(\alpha)L/(2r)\log\left[  1+4r^{2}\sin^{2}(\alpha)/L^{2}\right]  \}~.
\end{gather}

for $sin(\alpha)\neq0$. A simple calculation shows that when $L\rightarrow0$,
$F_{2}(L,r,r,\alpha)\rightarrow0$, regardless of $\alpha$ \ This is also true
when $d\neq r$, and $F_{2}$ is thus continuous when $L\rightarrow0$.

\subsubsection{Calculation of $F_{3}$}

From table \ref{tab:IntegralsFiDef},%

\begin{equation}
F_{3}(L,d,r,\alpha)=1/2-f_{2}(L,d,r,\alpha,\varphi)|_{\varphi_{\min}}%
^{\varphi_{\max}}~.\label{F3_bis_def}%
\end{equation}
The calculation of $f_{2}$ has already been done although the quantities
involved ($\phi_{+}$, $\rho_{+}$) have a different graphical interpretation as
shown in fig. \ref{fig9}. Eqs. \ref{phi_plus_def} and \ref{A21_integrate} can
still be used but some care is required as explained further on. From
eqs. \ref{F3_bis_def} and \ref{f2_A0_A1_A2} follows that%

\begin{equation}
F_{3}=\frac{1}{2\pi}\left[  A_{30}+\cos(\alpha)A_{31}+\sin(\alpha
)A_{32}\right]  _{\varphi_{\min}}^{\varphi_{\max}}~,\label{F3_A0_A1_A2}%
\end{equation}
where%

\begin{align}
A_{30}  & =\varphi-A_{20}~,\label{A30_def}\\
A_{31}  & =-A_{21}~,\label{A31_def}\\
A_{32}  & =-A_{22}\label{A32_def}%
\end{align}
and $A_{2i}$, $(i=1..3)$ are determined from eqs. \ref{A20_def},
\ref{A21_integrate} and \ref{A22_integrate}.

\subsubsection{Evaluation of $F_{3}$ ($d\neq0$)\label{F3_evaluation}}

With $\varphi_{\max}$ and $\varphi_{\min}$ chosen from table
\ref{tab:PhiLimits}%

\begin{align}
\rho_{+}(\varphi_{\max})  & =\widetilde{\rho}_{+}~,\\
\rho_{+}(\varphi_{\min})  & =-\widetilde{\rho}_{-}~,
\end{align}
where $\widetilde{\rho}_{-}$ and $\widetilde{\rho}_{+}$ are found from eqs.
\ref{rho_minus_tilde_def} and \ref{rho_plus_tilde_def}. To calculate
$A_{31}|_{\varphi_{\min}}^{\varphi_{\max}}$ the quantities%

\begin{equation}
a_{3}=\tan\left[  \frac{\phi_{+}}{2}\right]  _{\varphi_{\min}}=\frac
{\cos\left(  \alpha\right)  \widetilde{\rho}_{-}}{r-d+\sin\left(
\alpha\right)  \widetilde{\rho}_{-}}%
\end{equation}
and%

\begin{equation}
b_{3}=\tan\left[  \frac{\phi_{+}}{2}\right]  _{\varphi_{\max}}=\frac
{\cos\left(  \alpha\right)  \widetilde{\rho}_{+}}{r-d+\sin\left(
\alpha\right)  \widetilde{\rho}_{+}}%
\end{equation}
will be useful. $A_{30}$ is given by%

\begin{equation}
A_{30}|_{\varphi_{\min}}^{\varphi_{\max}}=\pi-\left[  \arctan\left(
L/\widetilde{\rho}_{+}\right)  -\arctan\left(  L/\widetilde{\rho}_{-}\right)
\right]  ~.
\end{equation}
Using eq. \ref{appendix_sum_of_ArcTan}\ from the appendix results in%

\begin{equation}
A_{30}|_{\varphi_{\min}}^{\varphi_{\max}}=\pi-2\arctan\left(  1/z\right)  ~,
\end{equation}
where%

\begin{equation}
z=\frac{r^{2}-d^{2}-L^{2}+\sqrt{\left(  r^{2}-d^{2}-L^{2}\right)  ^{2}%
+4L^{2}(r^{2}-\left[  d\cos(\alpha)\right]  ^{2})}}{2L\sqrt{r^{2}-\left[
d\cos(\alpha)\right]  ^{2}}}~.
\end{equation}
Then, applying $\arctan\left(  z\right)  +\arctan\left(  1/z\right)  =\pi/2$
yields%
\begin{equation}
A_{30}|_{\varphi_{\min}}^{\varphi_{\max}}=2\arctan\left(  z\right)  ~.
\end{equation}
Some care is required to obtain $A_{31}$. From fig. \ref{fig9} is clear that
as $\alpha\rightarrow$ $\pi/2$ \ then $\phi_{+}\left(  \varphi_{\min}\right)
$ $\rightarrow-\pi$ and that $-\pi<\phi_{+}\left(  \varphi_{\min}\right)
<-3\pi/2$ for $\alpha>$ $\pi/2$. Therefore $a_{3}=\tan\left[  \phi_{+}%
(\varphi_{\min})/2\right]  $ is not continuous as $\alpha$ goes through
$\pi/2$. The discontinuity also appears in the rhs of eq. \ref{phi_plus_def}
when $\varphi$ decreases to less than $-\pi$. From eqs. \ref{A31_def} and
\ref{A21_integrate} its obvious that both terms in $A_{31}$ will be
discontinuous. The integral, of course, must be continuous and a proper
expression can be provided by noticing that while%

\begin{equation}
z\neq\arctan\left[  \tan(z)\right]  \;;-\pi<z<-\pi/2~,
\end{equation}
one has%

\begin{equation}
z=-\pi/2-\arctan\left[  1/\tan\left(  z\right)  \right]  \;;-\pi<z<0~.
\end{equation}
Thus, to evaluate $A_{31}$, the substitutions%

\begin{equation}
\arctan(a_{3})\rightarrow-\pi/2-\arctan(1/a_{3})~,
\end{equation}

\begin{equation}
\arctan\left(  \sqrt{\frac{1-m}{1+m}}a_{3}\right)  \rightarrow-\pi
/2-\arctan\left[  1/\left(  \sqrt{\frac{1-m}{1+m}}a_{3}\right)  \right]  ~,
\end{equation}
are used and it is found that%

\begin{equation}
A_{31}|_{\varphi_{\min}}^{\varphi_{\max}}=-\frac{L}{2d}\left\{  2\left[
\pi/2+\Delta t_{1}\right]  -\frac{2}{\sqrt{1-m^{2}}}\left[  \pi/2+\Delta
t_{2}\right]  \right\}  ~,\label{A30_evaluation}%
\end{equation}
where%

\begin{equation}
\Delta t_{1}=\arctan(b_{3})+\arctan(1/a_{3})~,
\end{equation}

\begin{equation}
\Delta t_{2}=\arctan(\sqrt{\frac{1-m}{1+m}}b_{3})+\arctan(\sqrt{\frac
{1+m}{1-m}}/a_{3})~.
\end{equation}
The sum of arctangents can be cast as a single one using eq.
\ref{appendix_sum_of_ArcTan} from the appendix and after some algebra one gets%

\begin{equation}
\Delta t_{1}=2\arctan\left[  d\cos\left(  \alpha\right)  /\left(  r+E\right)
\right]  ~,
\end{equation}

\begin{equation}
\Delta t_{2}=2\arctan\left[  \left\{  Gd\cos\left(  \alpha\right)  \right\}
/\left\{  EH+r\sqrt{H^{2}-\left[  2Ld\cos\left(  \alpha\right)  \right]  ^{2}%
}\right\}  \right]  ~,
\end{equation}
where%

\begin{equation}
E=\sqrt{r^{2}-\left[  d\cos(\alpha)\right]  ^{2}}~,
\end{equation}

\begin{equation}
G=L^{2}+d^{2}-r^{2}%
\end{equation}
and%

\begin{equation}
H=\sqrt{\left[  L^{2}+\left(  d+r\right)  ^{2}\right]  \left[  L^{2}+\left(
d-r\right)  ^{2}\right]  }~.
\end{equation}
The remaining term is given by%

\begin{equation}
A_{32}|_{\varphi_{\min}}^{\varphi_{\max}}=-\frac{L}{2d}\log\left(  \frac
{L^{2}+\widetilde{\rho}_{+}^{2}}{L^{2}+\widetilde{\rho}_{-}^{2}}\right)  ~.
\end{equation}

\subsubsection{Evaluation of $F_{3}$ ($d=0$)}

Calculation of the limit of the expressions obtained in \ref{F3_evaluation}
yields $F_{3}(L,d=0,r,\alpha)=1/2-(\arctan\left[  L/r\right]  +Lr/\left(
L^{2}+r^{2}\right)  )$. In the same way one can show that $F_{3}%
(L\rightarrow0,d<r,r,\alpha)=1/2$ regardless of $d$, $r$ and $\alpha$. Altough
$F_{3}$ is only needed for $d<r$, it is worth mentioning that $\underset
{L\rightarrow0}{\lim}(\underset{d\rightarrow r-}{\lim}F_{3}(L,d,r,\alpha
))=1/4(1+\cos\alpha)$.

\section{Sample graphics and discussion}

\begin{figure}[h]
\begin{center}
\includegraphics[
height=5.0918cm,
width=7.8002cm
]{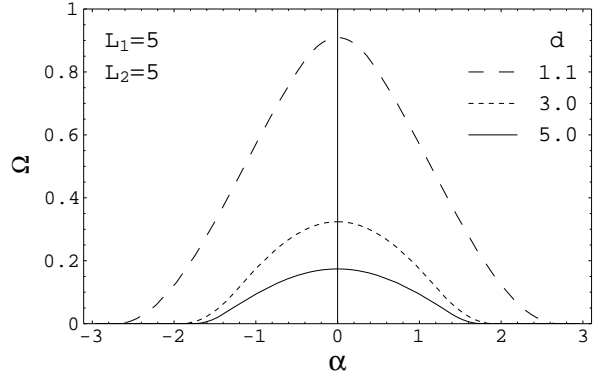}
\caption{Solid angle in case (i), for a cylinder with radius $1$ and length
$10=5+5$}
\label{fig10}
\end{center}
\end{figure}

\begin{figure}[h]
\begin{center}
\includegraphics[
height=5.0918cm,
width=8.1495cm
]{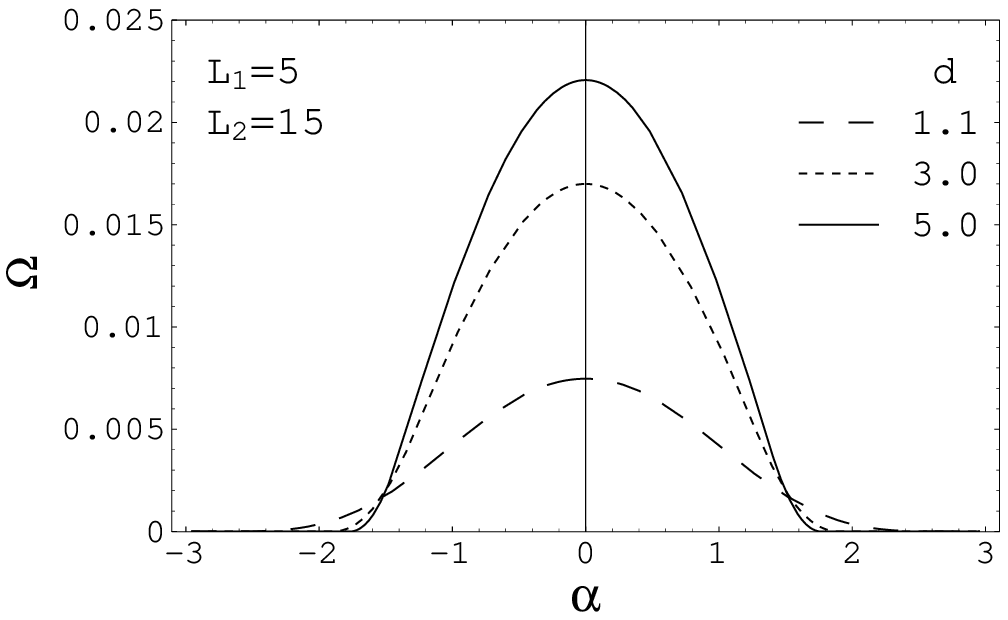}
\caption{Solid angle in case (ii) for a cylinder with radius $1$ and length
$10=15-5$}
\label{fig11}
\end{center}
\end{figure}

\begin{figure}
[ptb]
\begin{center}
\includegraphics[
height=5.0918cm,
width=8.3428cm
]{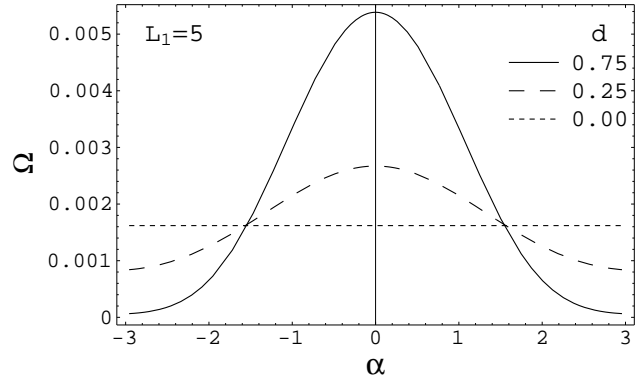}
\caption{Solid angle in case (iii) when $\Omega=\Omega_{circ}$, for a circle
with radius $1$. The intersection of the three curves near $\pi/2$ is not
exact.}
\label{fig12}
\end{center}
\end{figure}

\begin{figure}[h]
\begin{center}
\includegraphics[
height=5.0918cm,
width=8.1495cm
]{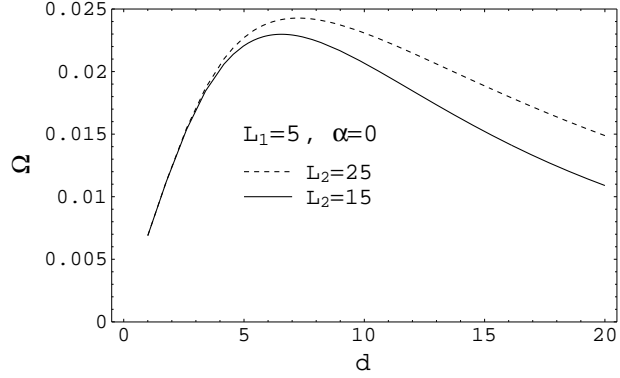}
\caption{Peak values ($\alpha=0$) of the solid angle in case (ii) as a
function of $d$, for two cylinders of radius $1$ and lengths $10$ ($15-5$) and
$20$ ($25-5$)}
\label{fig13}
\end{center}
\end{figure}

To illustrate the behavior with respect to $\alpha$ we consider a cylinder of
lenght 10 and radius 1. In figs. \ref{fig10}, \ref{fig11} and \ref{fig12} are
shown examples of cases (i),(ii) and (iii), respectively. As argued before,
$\Omega$ is an even function of $\alpha$ in all cases. In cases (i) and (ii),
for $0\leq\left\vert \alpha\right\vert <\pi/2-\arcsin(r/d)$, the solid angle
is \textit{simply proportional} \textit{to} $cos(\alpha)$; for $\left\vert
\alpha\right\vert >\pi/2+\arcsin(r/d)$, $\Omega=0$. In between, the dependence
on $\alpha$ is more complicated because a fraction of the cylinder is not
illuminated by the source, but $\Omega(\alpha)$ obviously decreases as
$\left\vert \alpha\right\vert $ increases. The transition region is large for
$d\approx r$ ($d<r$) and becomes increasingly narrower as $d$ increases. For
$d\gg r$, the region essentially vanishes so that $\Omega\varpropto
cos(\alpha)$ when $\left\vert \alpha\right\vert \lessapprox\pi/2$ and
$\Omega=0$ when $\left\vert \alpha\right\vert \gtrapprox\pi/2$. It is
intuitive that in case (i), $\Omega$ decreases when distance $d$ increases and
all other parameters are held constant. This is not necessarily true for case
(ii), because source and cylinder are in a skew geometry. Considering two
distances $d_{1}<d_{2}$, one concludes from the preceding discussion that
there is certainly some region for $\alpha$ where $\Omega(d_{2})=0$ and
$\Omega(d_{1})>0$ so that $\Omega(d_{1})>\Omega(d_{2})$. As $\left\vert
\alpha\right\vert $ decreases to $0$, this relation \textit{can} be inverted,
which happens for all distances shown in fig. \ref{fig11}. Since for very
large $d$, $\Omega$ is certainly a decreasing function of $d$, there must be a
maximum, for each $\alpha.$ This is illustrated in fig. \ref{fig13}, for
$\alpha=0$, $L_{1}=5$ and two cylinder lengths: 10 ($L_{2}=15$) and 20
($L_{2}=25$). It is worth mentioning that in case (ii), apart form the
$\alpha$ dependence, a similar behavior can be observed for an isotropic
point source (i.e. a maximum when $d$ is changed and $L_{1}$ and $L_{2}$ kept
constant). Case (iii) is exemplified in fig. \ref{fig12}.\ In this case
($r>d$) the solid angle is defined only by one the end circles which is, in
general, partially illuminated by the source. The exception happens when
$d\longrightarrow r$ : for $\alpha=0$ the circle is fully illuminated where as
for $\alpha=\pi$ it is totally obscured. When $d=0$ the source is aligned with
the center of circle and $\Omega$ is then independent of $\alpha$. As $d$
increases, for $L_{1}\neq0$, the dependence on $\alpha$ becomes stronger with a
maximum when $\alpha=0$. When $L_{1}=0$ (not shown in fig. \ref{fig12}),
$\Omega=1/2$, regardless of $\alpha$ or $d$.%

\pagebreak

\section{Conclusions and outlook}

The solid angle defined by a point cosine source and a right circular cylinder
with axis orthogonal to that of the source has been treated analitically. It
has been shown that for $d>r$ and $0\leqslant\left\vert \alpha\right\vert
<\pi/2-\arcsin(r/d)$, the whole dependence on $\alpha$ is given by a
$cos(\alpha)$ factor (eqs. \ref{F1_even} and \ref{F2_even}). It is possible to
obtain, in a similar fashion, analytical expressions in the case of a cosine
source distributed on a wire paralell to the cylinder axis. A work where we
report these results is in preparation.

\begin{ack}
I would like to thank Jo\~{a}o Prata
for his thorough review of the manuscript.
\end{ack}

\appendix

\section{Sum of Arctans}

\label{Appendix} To prove the identity
\begin{equation}
\arctan x\pm\arctan y=2\arctan\left[  \left(  x\pm y\right)  /\left(  1\mp
xy+\sqrt{\left(  x\pm y\right)  ^{2}+\left(  1\mp xy\right)  ^{2}}\right)
\right] \label{appendix_sum_of_ArcTan}%
\end{equation}
consider $h(x,y)=\arctan x\pm\arctan y$ and $g(x,y)=2\arctan\left[  \left(
x\pm y\right)  /D\right]  $ where
\[
D=\left(  1\mp xy+\sqrt{\left(  x\pm y\right)  ^{2}+\left(  1\mp xy\right)
^{2}}\right)  ~.
\]
Using $\tan(u\pm v)=\left[  \tan(u)\pm\tan(v)\right]  /\left[  1\mp\tan
(u)\tan(v)\right]  $, we find that $\tan\left(  h\right)  =\left(  x\pm
y\right)  /\left(  1\mp xy\right)  $. In a similar way, applying
$\tan(2u)=2\tan(u)/\left[  1-\tan^{2}(u)\right]  $, gives%

\begin{align*}
\tan\left[  g(x,y)\right]    & =2\frac{\left(  x\pm y\right)  /D}{\left[
D^{2}-\left(  x\pm y\right)  ^{2}\right]  /D^{2}}\\
& =\left(  x\pm y\right)  /\left(  1\mp xy\right)  ~,
\end{align*}
where in the last step we used the fact that $\ D\neq0$. Since $\tan\left(
h\right)  =\tan\left(  g\right)  $ then $h-g=n\pi$ for some n. Taking $x=y=0$
results that $n=0$ and, because $h$ and $g$ are continuous, we conclude that
eq. \ref{appendix_sum_of_ArcTan} holds for any $x$ and $y$.

\end{document}